\documentclass[a4paper]{cas-dc}

\usepackage[numbers,sort&compress]{natbib}

\def\tsc#1{\csdef{#1}{\textsc{\lowercase{#1}}\xspace}}
\tsc{WGM}
\tsc{QE}
\tsc{EP}
\tsc{PMS}
\tsc{BEC}
\tsc{DE}
\usepackage[title]{appendix} % for correct naming of appendices
\usepackage{ulem} %for crossing/cancelling text with \sout{} & \xout{}
\usepackage{float}
\usepackage{subcaption}
\usepackage{longtable} %for long tables spanning on several pages
\DeclareMathAlphabet\mathbfcal{OMS}{cmsy}{b}{n}
\newcommand{\comentar}[1]{}

\newcommand{\JLBG}[1]{{\leavevmode\color{cyan}#1}}

\newcommand{\tensa}[1]{\overset{\lower.5em\hbox{\text{\tiny$\leftrightarrow$}}}{\mathbf{#1}}}
\newcommand{\tensb}[1]{\overset{\lower.5em\hbox{\text{\tiny$\leftrightarrow$}}}{\pmb{#1}}}
\newcommand{\vv}[1]{\vec{\mathbf{#1}}}
\newcommand{\vvb}[1]{\vec{\pmb{#1}}}
\renewcommand{\arraystretch}{3.0}
\newcommand{\var}[1]{\left( #1 \right) }

\newcommand{\abs}[1]{\left| #1 \right| }
\newcommand{\w}[0]{\omega }
\newcommand{\EE}[0]{\vec{\mathbf{E}}_{\text{ext}}}

\newcommand{\ee}[1] {\hat{\mathbf{e}}_{#1}}
\newcommand{\wbgv}[0]{ \left( \frac{\left|\omega\right|b}{\gamma v} \right) }

\usepackage[pangram]{blindtext}
\usepackage{lipsum}

\graphicspath{{figs/}} %Setting the graphicspath

\begin{document}
		
\let\WriteBookmarks\relax
\def\floatpagepagefraction{1}
\def\textpagefraction{.001}
\shorttitle{Angular momentum transfer from swift electrons to non-spherical nanoparticles in the dipolar approximation}
\shortauthors{J. L. Briseño-Gómez et~al.}

\title [mode = title]{Angular momentum transfer from swift electrons to non-spherical nanoparticles within the dipolar approximation}     

\author[1]{Jorge Luis Brise\~no-G\'omez}[orcid=0000-0001-8412-4868]
\cormark[1]
\ead{jorgeluisbrisenio@ciencias.unam.mx}

\address[1]{Departamento de F\'isica, Facultad de Ciencias, Universidad Nacional Aut\'onoma de M\'exico, Ciudad Universitaria, Av. Universidad \#3000, Ciudad de M\'exico 04510, Mexico.}

\address[2]{Department of Physics and Astronomy, Uppsala University, Box 516, 75120 Uppsala, Sweden.}

\author[1]{Atzin L\'opez-Tercero}

\author[2]{Jos\'e {\'A}ngel Castellanos-Reyes}

\author[1]{Alejandro Reyes-Coronado}

\cortext[cor1]{Corresponding author}

%************************************************************
%************************************************************
%								ABSTRACT
%************************************************************
%************************************************************
\begin{abstract}
 In this work, we study the angular momentum transfer from a single swift electron to non-spherical metallic nanoparticles, specifically investigating spheroidal and polyhedral (Platonic Solids) shapes. While previous research has predominantly focused on spherical nanoparticles, our work expands the knowledge by exploring various geometries. Employing classical electrodynamics and the small particle limit, we calculate the angular momentum transfer by integrating the spectral density, ensuring causality through Fourier-transform analysis. Our findings demonstrate that prolate spheroidal nanoparticles exhibit a single blueshifted plasmonic resonance, compared to spherical nanoparticles of equivalent volume, resulting in lower angular momentum transfer. Conversely, oblate nanoparticles display two resonances—one blueshifted and one redshifted—resulting in a higher angular momentum transfer than their spherical counterparts. Additionally, Platonic Solids with fewer faces exhibit significant redshifts in plasmonic resonances, leading to higher angular momentum transfer due to edge effects. We also observe resonances and angular momentum transfers with similar characteristics in specific pairs of Platonic Solids, known as duals. These results highlight promising applications, particularly in electron tweezers technology.
\end{abstract}

\maketitle

%************************************************************
%************************************************************
%								INTRODUCTION
%************************************************************
%************************************************************
\section{Introduction}

The study of electron-matter interactions has gained significant importance in research, driven in part by the widespread use of electron microscopy techniques to investigate the electronic and optical properties of materials \cite{GarciadeAbajo-1, kirkland1998advanced, carter2016transmission}. Electron microscopy studies have become standard practice, offering atomic-level spatial resolution and ultrafast femtosecond temporal scales \cite{barwick2009, Krivanek1, Krivanek2, pennycook2012seeing, dabrowski2020ultrafast, egerton2011electron}. Additionally, experimental evidence has demonstrated the ability to manipulate samples using electron beams \cite{Oleshko, verbeeck2013, susi2017manipulating}.

Electron energy loss spectroscopy (EELS) is a widely utilized technique in electron microscopy, crucial for investigating optical excitations in materials \cite{GarciadeAbajo-1}. Recent advances in electron vortex technologies have given rise to vortex-EELS, enabling precise mapping of the local magnetic responses of materials at the atomic scale \cite{mohammadi2012, Lloyd}. Electron vortices, characterized by specific orbital angular momentum, can transfer their angular momentum to nanoparticles (NPs) due to their distinct mechanical and electromagnetic properties \cite{lloyd2013}. However, experimental studies with gold NPs have shown discrepancies with theoretical predictions that overestimate the angular momentum transfer by three orders of magnitude, highlighting the complexities involved in these interactions \cite{verbeeck2013}.

In the dynamic landscape of electron microscopy, novel techniques have emerged, like 4D-EELS, offering simultaneous measurement of energy and momentum transfers, thereby unveiling profound insights into nanoscale material properties \cite{zewail2010four}. Particularly groundbreaking is the capacity for single-atom manipulation, realized through meticulous control of momentum transfer from electrons. Advanced a\-be\-rra\-tion-corrected instruments, capable of varying electron energy while preserving resolution, empower researchers to shape materials at the atomic level, achieving a significant milestone in materials science and engineering \cite{susi2017manipulating}.

In previous theoretical and experimental research, electron beams without vorticity were shown to transfer both linear and angular momentum to nanoparticles, describing the interaction through classical electrodynamics \cite{Batson01,Batson0,GarciadeAbajo0,Batson,PRBCoronado,Vesseur,Batson2,Batson3,Lagos,Lagos2,castellanos2019,castellanos2023theory}. Moreover, many of the previous theoretical studies focused primarily on the angular momentum transfer to spherical nanoparticles. In particular, an investigation of the angular dynamics induced on small spherical nanoparticles by electron beams was conducted, applying a classical electrodynamics approach within the small particle limit \cite{castellanos2021angular}. This approach considered the interaction with a single swift electron, justified by the considerably lower time of electromagnetic excitations inside metals compared to the emission time of individual electrons within typical Scanning Transmission Electron Microscope (STEM) electric currents. The previous study provided expressions for the torque and angular momentum transferred from a swift electron to a small NP, based on its polarizability. This foundation allows for the extension of the analysis to various geometries beyond spheres.

In this work, we delve into the angular momentum transfer to nonspherical NPs, focusing on the influence of nanoparticle geometry. Using analytically derived polarizabilities for spheroidal NPs \cite{bohren} and numerically computed polarizabilities for Platonic Solids \cite{sihvola2004polarizabilities}, we study the angular momentum transfer to aluminum and gold NPs, examining the relationship between the magnitude of the angular momentum transfer and plasmonic resonances. It is worth mentioning that, despite the apparent geometric complexity of Platonic Solids compared to spheres, their polarizability is a scalar quantity owing to their symmetries. 

The insights gathered from our study illuminate the intricate dynamics of angular momentum transfer in these systems, offering valuable guidance for potential applications.

%************************************************************
%************************************************************
%							METHOD
%************************************************************
%************************************************************
\section{Theoretical model for the angular momentum transfer from a swift electron to an NP in the dipole approximation}

\begin{figure}[h!]
 \centering
\includegraphics[width=0.65\linewidth]{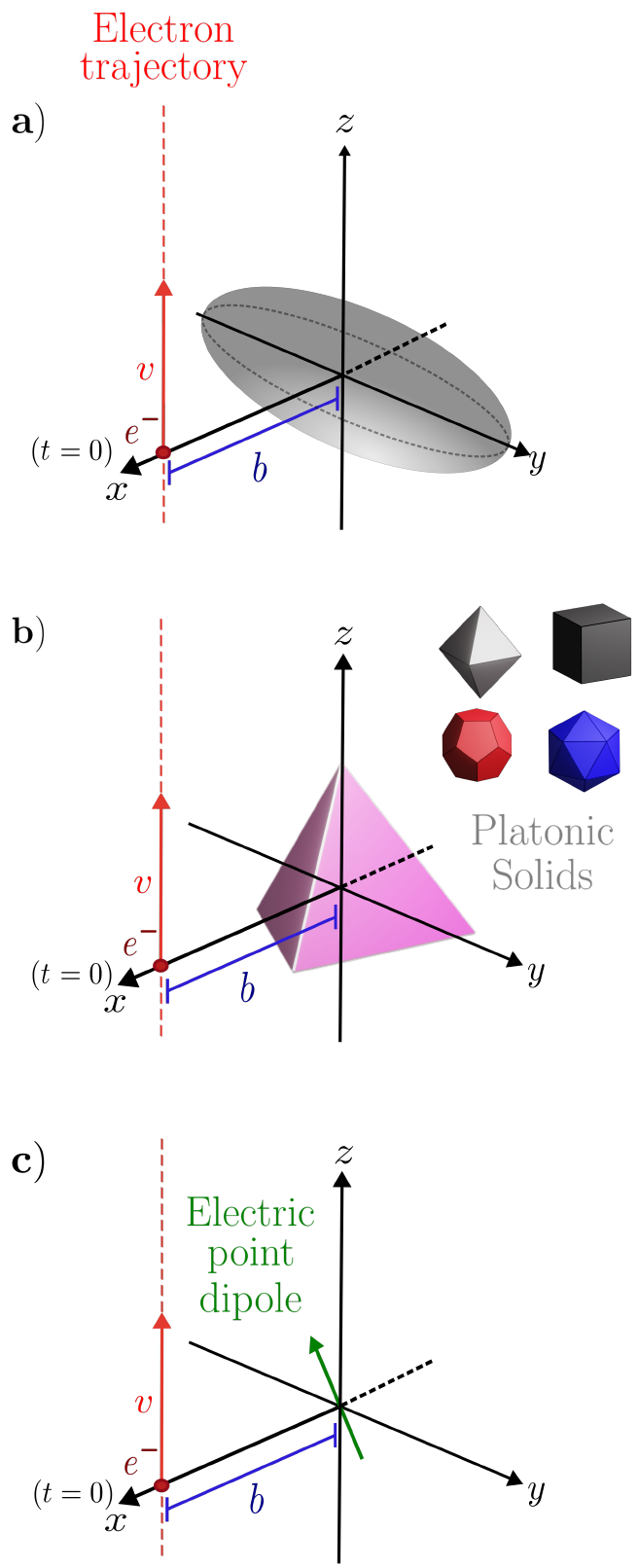}
\caption{\label{fig: system} Diagrams showing an electron traveling with constant speed in the neighborhood of the different non-spherical nanoparticles considered in this work: \textbf{a)} an ellipsoid (see Table \ref{tab: Ellipsoids}), \textbf{b)} the Platonic solids (see Table \ref{tab: Platonic Solids}), and \textbf{c)} an electric point dipole}.
\end{figure}

In this section, we present an overview of the formalism \cite{castellanos2021angular} regarding the interaction between a swift electron and a small nanoparticle (NP). The trajectory of the electron is defined by $\vec{r}(t) = (b, 0, vt)$, where $b$ represents the impact parameter and $v$ is its constant velocity. The origin of the coordinate system is positioned inside the NP, as depicted in Figure \ref{fig: system}. This model assumes a time range from $-\infty$ to $\infty$, with the swift electron being at a distance $b$ from the origin at $t = 0$. The NP consists of a homogeneous material with an electromagnetic response characterized by the dielectric function $\epsilon(\omega)$. For our calculations, we utilize the bulk dielectric function $\epsilon(\omega)$, since it has been demonstrated to be adequate to compute angular momentum transfers to spherical NPs with radii $\geq 1$~nm \cite{castellanos2023theory}. 

We investigate the interaction between a single swift electron and either a spheroidal (prolate or oblate) NP, as shown in Figure \ref{fig: system}$\mathbf{a)}$, or one shaped as a Platonic Solid, illustrated in Figure \ref{fig: system}$\mathbf{b)}$. Within the dipolar approximation, the electromagnetic response of the NP is assumed to be that of an electric point dipole positioned at the origin, as depicted in Figure \ref{fig: system}$\mathbf{c)}$.

For the discussion, SI units will be employed. In the small particle approximation, the traveling electron generates an external electric field $\mathbf{E}(\mathbf{r}, \omega)$, inducing an electric dipole moment within the NP, presumed to be composed of a nonpolar material:

\begin{equation}
\vv{p}\var{\vv{r},\w} = \epsilon_0 \, \tensa{\alpha}\var{\w}\cdot\EE\var{\vv{r},\w},
\end{equation}
where $\epsilon_0$ denotes the vacuum permittivity, and $\tensa{\alpha}\var{\w}$ represents the electric polarizability tensor characterizing the electromagnetic response of the NP. The angular momentum transfer (AMT) can be expressed in terms of $\tensa{\alpha}\var{\w}$ and $\omega$ as follows \cite{castellanos2021angular}:

\begin{equation}
\Delta \vv{L} = \int_0^{\infty} \vvb{\mathcal{L}}(\tensa{\alpha}; \omega) d\omega,
\label{eq: DL integral spectral density}
\end{equation}
where $\vvb{\mathcal{L}}(\tensa{\alpha}; \omega)$ is referred to as the \textit{spectral density} of the angular momentum transfer. The complete expressions for the components of the spectral densities in the general non-symmetric case are provided in Appendix \ref{Appendix: Spectral densities}.
 
The analytic expression for the spectral density in the case of small homogeneous NPs with diagonal polarizability tensor, $\tensa{\alpha}=\alpha_{\rm x}\ee{x}\ee{x}+\alpha_{\rm y}\ee{y}\ee{y}+\alpha_{\rm z}\ee{z}\ee{z}$, is 
\begin{align}
    \vvb{\mathcal{L}}\var{\tensa{\alpha}; \w}=& -\var{\frac{1}{4\pi\epsilon_0}}^2 \frac{4 q_e^2}{v^4 \gamma^2}\frac{\epsilon_0}{\pi} \bigg[ \frac{\abs{\w}\w}{\gamma} K_0\wbgv \times \nonumber\\
    & K_1\wbgv \text{Im} \Big\{ \alpha_{\rm x}\var{\w} + \alpha_{\rm z}\var{\w} \Big\}\bigg] \ee{y},
    \label{eq: DLy diagonal}
\end{align}
whereas the analytical expression for the spectral density in the case of a small homogeneous and isotropic NP characterized with a scalar electric polarizability, $\alpha\var{\w}$, is given by  
\begin{align}
\vvb{\mathcal{L}}\var{\alpha; \w} =& -\var{\frac{1}{4\pi\epsilon_0}}^2\frac{8 q_e^2}{v^4 \gamma^3}\frac{\epsilon_0}{\pi} \bigg[\w\abs{\w} K_0\var{\frac{\w b}{\gamma v}} \times \nonumber \\ & K_1\var{\frac{\w b}{\gamma v}} \bigg] {\rm Im}\big\{\alpha\var{\w}\big\} \ee{y},
\label{eq: DLy non-diagonal}
\end{align}
where $-q_e$ is the electric charge of the electron, $K_{\nu}$ is the modified Bessel function of the second kind of order $\nu$, $v$ is the constant electron speed, $\gamma = (1-\beta^2)^{-1/2}$ is the Lorentz factor, with $\beta = v/c$, $c$ is the speed of light, and $\text{Im} \{z\}$ denotes the imaginary part of $z$.

For a small spherical NP, the polarizability is given by \cite{bohren}
\begin{equation}
\alpha(\omega) = 3V \frac{\epsilon(\omega)-\epsilon_{0}}{\epsilon(\omega)+2\epsilon_{0}} = 3V \frac{\tilde{\epsilon}(\w)-1}{\tilde{\epsilon}(\w)+2},
\end{equation}
where $V$ denotes the volume of the sphere and $\tilde{\epsilon}(\w) = \epsilon(\omega)/\epsilon_{0}$ is the relative dielectric function. From here on, to simplify the notation, we omit the $\omega$ dependence in the dielectric function.

%************************************************************
%************************************************************
%								ELLIPSOIDS
%************************************************************
%************************************************************
\subsection{Ellipsoids}
For small ellipsoids, analytical expressions of the polarizabilities are known \cite{bohren}. Given a specific axis, denoted by $i \in \{ x,y,z \}$, the polarizability of an ellipsoid is given by
\begin{equation}
\alpha_{i} = 3V \frac{\tilde{\epsilon}-1}{3 + 3 L_{i} (\tilde{\epsilon} - 1)},
\label{eq: alpha_spheroids}
\end{equation}
where $V$ represents the volume of the ellipsoid. The polarizability tensor is direction-dependent and corresponds to a diagonal tensor with three distinctive polarizabilities ($\alpha_x, \alpha_y, \alpha_z$). The geometric factors, denoted by $L_i$, are intricately linked to the depolarization factors. Detailed information regarding these factors can be found in Appendix \ref{Appendix: spheroids}.

\renewcommand{\arraystretch}{5.5}
\begin{table}[ht!]
\begin{tabular}{@{}ccc@{}}
\toprule
Figure   & Example & Characteristics \\ \midrule
\rowcolor[HTML]{F8F8F8} 
Generic ellipsoid  & \parbox[c]{8.em}{\vspace{-2.0em}\includegraphics[width=1\linewidth]{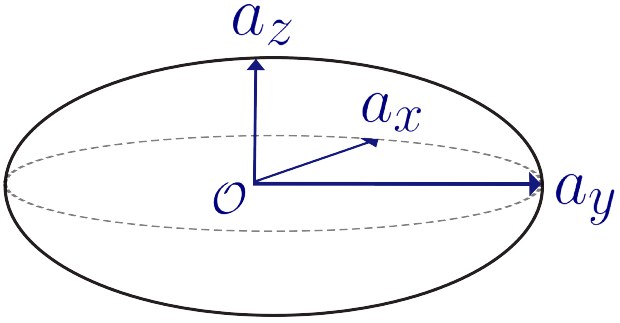}} & $a_z \le a_x \le a_y$ \\
Prolate spheroid   &  \parbox[c]{8.em}{\vspace{-2.0 em}\includegraphics[width=\linewidth]{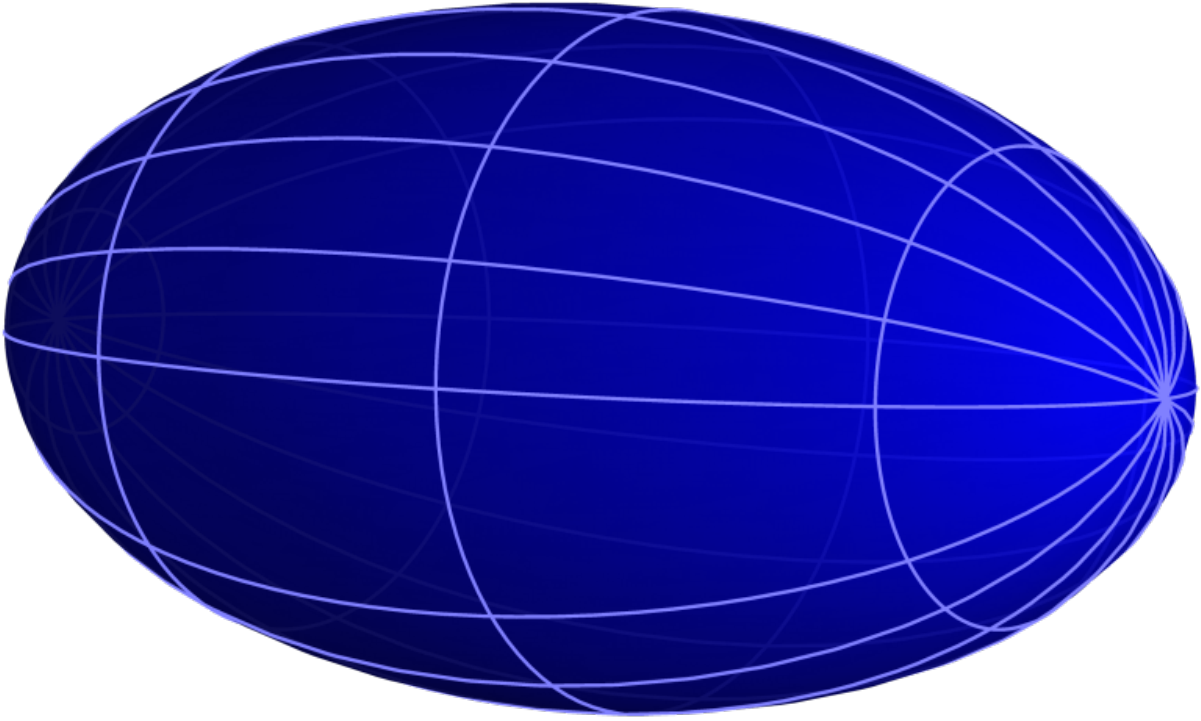}} & $a_z = a_x \leq a_y$ \\
\rowcolor[HTML]{F8F8F8} 
Oblate spheroid  &   \parbox[c]{8.5em}{\vspace{-2.0em}\includegraphics[width=1\linewidth]{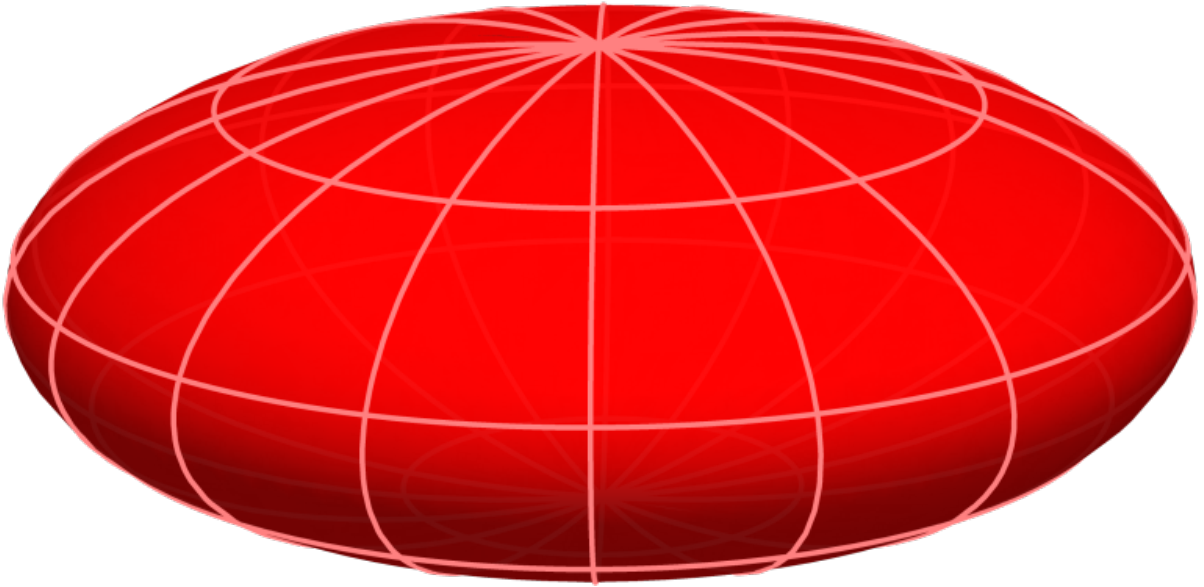}}   & $a_y = a_x \geq a_z$    \\ \bottomrule
\end{tabular}
\caption{\label{tab: Ellipsoids} List of the spheroids. Each entry includes the type of spheroid, a visual representation, and the characteristics that the semi-axes satisfy.} 
\end{table}

 In this study, we focus on the particular class of ellipsoids, known as spheroids, which possess two semiaxes of equal length. The characteristics of different types of spheroids are summarized in Table \ref{tab: Ellipsoids}, where each entry includes the type of spheroid, a visual representation, and the characteristics that the semi-axes satisfy. The prolate spheroid, satisfying the condition $a_z = a_x \leq a_y$, exhibits the same behavior in its plasmonic modes, the longitudinal mode (along $\ee{z}$) and the transverse mode (along $\ee{x}$), as detailed in Section \ref{sec: amt ellipsoids}. Conversely, the oblate spheroid, also outlined in Table \ref{tab: Ellipsoids}, satisfies $a_y = a_x \geq a_z$, resulting in an asymmetry between the longitudinal and transverse modes. This asymmetry leads to the splitting of these modes into distinct resonances, as further discussed in Section \ref{sec: amt ellipsoids}.

%************************************************************
%************************************************************
%								PLATONIC SOLIDS
%************************************************************
%************************************************************
\subsection{Platonic Solids}
\label{sec: Platonic Solids}
The Platonic Solids, presented in Table \ref{tab: Platonic Solids}, are the only five regular polyhedra with identical regular polygons as faces. They consist of the tetrahedron (or regular pyramid), hexahedron (or cube), octahedron, icosahedron, and dodecahedron. The polarizabilities of Platonic Solids have been determined in Ref. \cite{sihvola2004polarizabilities} through numerical computations, followed by analytical representations using Padé approximants. The derived analytical formula for each Platonic Solid can be expressed as:
\begin{equation}
\alpha(\omega) = V(\tilde{\epsilon} - 1) \frac{P(\tilde{\epsilon})}{Q(\tilde{\epsilon})},
\label{eq: polarizability PS}
\end{equation}
where $V$ represents the volume of the Platonic Solid, and $P(x)$ and $Q(x)$ are polynomials given in Ref. \cite{sihvola2004polarizabilities}. The full expressions of the polarizabilities of Platonic Solids can be found in Appendix \ref{Appendix: Polarizabilities}.

\renewcommand{\arraystretch}{4.5}
\begin{table}[hb!]
\begin{tabular}{@{}cccc@{}}
\toprule
Polyhedron   & Figure                                                                                                                                     & Faces & Dual polyhedron \\ \midrule
\rowcolor[HTML]{F8F8F8} 
Tetrahedron  & \parbox[c]{4.0em}{\vspace{-1.5em}\includegraphics[width=1\linewidth]{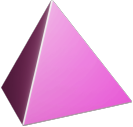}} & 4     &      \parbox[c]{4.0em}{\vspace{-1.5em}\includegraphics[width=1\linewidth]{3-Table2a-tetrahedron.pdf}     }      \\
Hexahedron   &  \parbox[c]{4.0em}{\vspace{-1.5em}\includegraphics[width=1\linewidth]{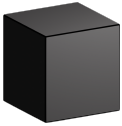}} & 6     &      \parbox[c]{4.0em}{\vspace{-1.5em}\includegraphics[width=1\linewidth]{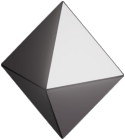}}           \\
\rowcolor[HTML]{F8F8F8} 
Octahedron   &   \parbox[c]{4.0em}{\vspace{-1.5em}\includegraphics[width=1\linewidth]{3-Table2c-octahedron.pdf}}   & 8     &   \parbox[c]{4.0em}{\vspace{-1.5em}\includegraphics[width=1\linewidth]{3-Table2b-hexahedron.pdf}}  \\
Dodecahedron &  \parbox[c]{4.0em}{\vspace{-1.5em}\includegraphics[width=1\linewidth]{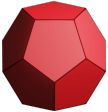}}  & 12    &  \parbox[c]{4.0em}{\vspace{-1.5em}\includegraphics[width=1\linewidth]{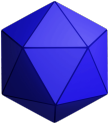}} \\ 
\rowcolor[HTML]{F8F8F8} 
Icosahedron  &   \parbox[c]{4.0em}{\vspace{-1.5em}\includegraphics[width=1\linewidth]{3-Table2e-icosahedron.pdf}} & 20    &  \parbox[c]{4.0em}{\vspace{-1.5em}\includegraphics[width=1\linewidth]{3-Table2d-dodecahedron.pdf}} \\ \bottomrule
\end{tabular}
\caption{\label{tab: Platonic Solids} List of the Platonic Solids along with their corresponding dual polyhedron. Each entry includes the name of the polyhedron, a visual representation, the number of faces in the polyhedron, and the visual representation of its dual polyhedron.}
\end{table}

Dual Platonic Solids (which can be constructed through a process of reciprocation \cite{ede1984dual}, resulting in an exchange of faces and vertices while maintaining the same number of edges) exhibit a closely linked electromagnetic response, related to a reduction in the number of faces \cite{sihvola2004polarizabilities}. This characteristic leads to a similar angular momentum transfer, as discussed later in this work. It is worth mentioning that the dual of a Platonic Solid is another Platonic Solid, as illustrated in Table \ref{tab: Platonic Solids}. This characteristic lies at the heart of the interconnected electromagnetic properties between dual Platonic solids, constituting a fundamental aspect of our analysis.

%************************************************************
%************************************************************
%						RESULTS AND DISCUSSION
%************************************************************
%************************************************************
\section{Results and discussion}

%Deben ser cortos.

%Deben ser organizados.

%Explicar los resultados y discutir cómo ayudan al problema mencionado anteriormente. Resumir los resultados, discutir si los resultados son esperados o inesperados, interpretar y explicar los resultados, hacer hipótesis sobre su generalidad.

%Debe de ir de lo particular (resultados en el trabajo) a lo general (cómo estos resultados demuestran un principio general más amplio). Discutir problemas encontrados.}

The calculation of the angular momentum transfer (AMT) using Eq. \eqref{eq: DL integral spectral density} relies on the election of the polarizability $\alpha(\epsilon)$ and, thus, the dielectric function $\epsilon(\omega)$ of the target material. Careful consideration must be exercised in this selection process, as prior studies have revealed the impact of non-causal electromagnetic responses, which can yield non-physical momentum transfers  \cite{castrejon2021effects}.

In this section, we outline the criteria for the choice of appropriate polarizabilities and present the insights gathered from computing the AMT using Eq. \eqref{eq: DL integral spectral density}. Furthermore, we conduct a comparative analysis of the AMT across different types of NPs, accounting for their respective spectral density resonances. For the sake of fairness in the comparison, we maintain the same volume for every NP considered in this study. Specifically, the volume of each NP is  the same as that of a sphere with a radius of $a=1$ nm ($V= 4\pi a^3/3$). This standardization allows for a meaningful examination of the AMT while controlling for variations in NP size and shape.

\subsection{Causality of the polarizabilities}

Investigating the causality of electromagnetic responses is critical in studying momentum transfers from swift electrons to NPs. A previous study \cite{castrejon2021effects} highlighted the potential appearance of non-physical outcomes arising from non-causal electromagnetic responses. Therefore, it is imperative to establish the causality of the polarizabilities governing the electromagnetic response of the NPs under examination.

\begin{figure}[h!]
\centering
\includegraphics[width=1\linewidth]{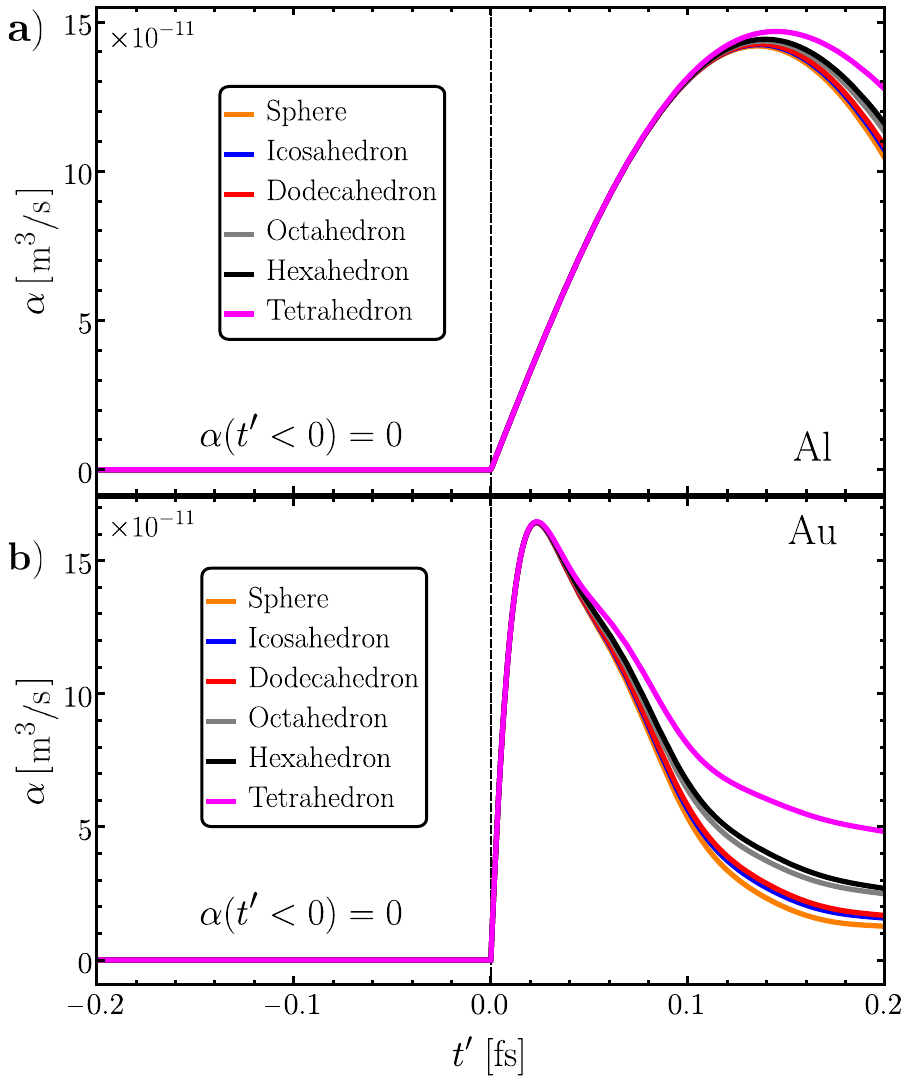}
\caption{\label{fig: polCausality} 
Polarizabilities as a function of time for different Platonic Solids and the sphere considering two materials: $\mathbf{a)}$ aluminum modeled with the Drude model \cite{rakic}, and $\mathbf{b)}$ gold from Werner et al. \cite{werner}. The polarizabilities exhibit a zero response for $t^{\prime} < 0$, indicating a causal behavior of each system.}
\end{figure}

In Figure \ref{fig: polCausality}, we present the polarizability of nanoparticles shaped as Platonic Solids as a function of time, computed through a Fourier Transform of Eq. \eqref{eq: polarizability PS}, for two distinct dielectric functions: Figure \ref{fig: polCausality}$\mathbf{a)}$ shows aluminum characterized by a Drude dielectric function with parameters $\hbar \omega_p = 13.142$ eV and $\hbar \Gamma = 0.197$ eV \cite{rakic}, while Figure \ref{fig: polCausality}$\mathbf{b)}$ illustrates gold as reported by Werner et al. \cite{werner}. The analysis encompasses geometries of both spheres and Platonic Solids, revealing that the response for $t^{\prime} < 0$ is zero, thereby demonstrating the causal behavior of each system\footnote{It is worth mentioning that $t^{\prime}$ denotes the response time of electromagnetic excitations within the NP, which differs from the time $t$ associated with the trajectory of the electron}. The same is also true for the polarizabilities of the spheroidal NPs, as shown in Appendix \ref{Appendix: spheroids}.

These results reinforce and verify the validity and reliability of the selected polarizabilities for subsequent calculations. 

\subsection{Angular momentum transfer for spheroids}
\label{sec: amt ellipsoids}

Equations \eqref{eq: DLy diagonal} and \eqref{eq: DLy non-diagonal} imply that, due to the inherent symmetry in this problem, the angular momentum transfer occurs only along a direction parallel to the $y$-axis, as defined in the coordinate system depicted in Fig. \ref{fig: system}. Therefore, from now on, we will focus on the magnitude $\Delta L$ of the angular momentum transfer in this direction.

We first consider the relative difference $(\Delta L_p - \Delta L_s) / \Delta L_s$ between the AMT to Platonic solids ($\Delta L_p$, with $p$ representing the solid) and spheres ($\Delta L_s$).

\begin{figure}[h!]
    \centering
    \includegraphics[width=1\linewidth]{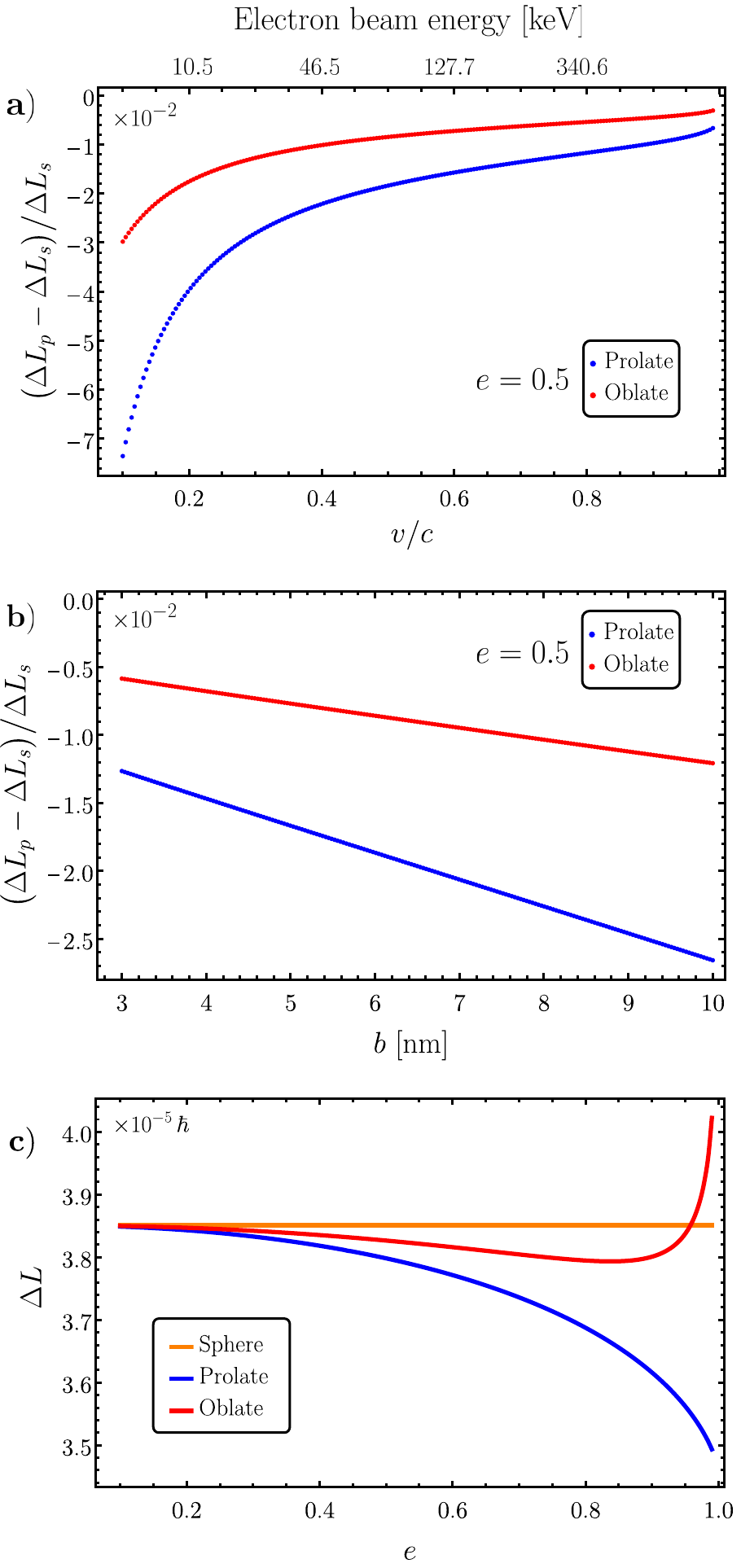}
    \caption{Relative difference in angular momentum transfer to a small ellipsoidal aluminum NP as a function of $\mathbf{a)}$ the speed of the electron $v$ with impact parameter $b = 3.5 \text{nm}$, and $\mathbf{b)}$ the impact parameter $b$ with $v = 0.7c$ ($200$ keV), for both prolate (blue) and oblate (red) ellipsoids. The eccentricity of the ellipsoids was set to $e = 0.5$. $\mathbf{c)}$ Angular momentum transfer to a small ellipsoidal aluminum NP as a function of the eccentricity, with $b = 3.5 \text{nm}$ and $v = 0.7c$ ($200$ keV).}
    \label{fig: LToSphereEllipsoids}
\end{figure}
In Figure \ref{fig: LToSphereEllipsoids}, we show through the plots of this relative difference that the AMT for a prolate spheroid (in blue) exceeds that of an oblate spheroid (in red). The relative difference in AMT is depicted as a function of the electron speed $v$ with $b=3.5 \text{nm}$ [see Fig. \ref{fig: LToSphereEllipsoids}$\mathbf{a)}$], and as a function of the impact parameter $b$ with $v=0.7 c$ (corresponding to a $200$ keV electron beam) [see Fig. \ref{fig: LToSphereEllipsoids}$\mathbf{b)}$], considering Al ellipsoids with an eccentricity $e = 0.5$. The enhanced response of prolate NPs can be attributed to the spectral characteristics of the electromagnetic field produced by a swift electron, which exhibits a more intense contribution at lower frequencies (further details can be found in Appendix \ref{Appendix: Ext Field}). Similar behavior is observed for the AMT to Au spheroidal NPs. A comparison between the results presented here and those for Au spheroidal nanoparticles in Appendix \ref{Appendix: Au NPs} offers insights into this consistency (compare Fig. \ref{fig: LToSphereEllipsoids} with Fig. \ref{fig: LToSphereEllipsoids-WernerAu}). Notably, depending on the values of eccentricity $e$, the AMT can surpass that of spherical particles, particularly for oblate nanoparticles with high eccentricity values ($e>0.95$). This trend is illustrated in Fig. \ref{fig: LToSphereEllipsoids}$\mathbf{c)}$, where the AMT for each ellipsoid is plotted as a function of $e$, with $b = 3.5$ $\text{nm}$ and $v = 0.7c$ (200 keV).

\begin{figure}[!h]
    \centering
    \includegraphics[width=1\linewidth]{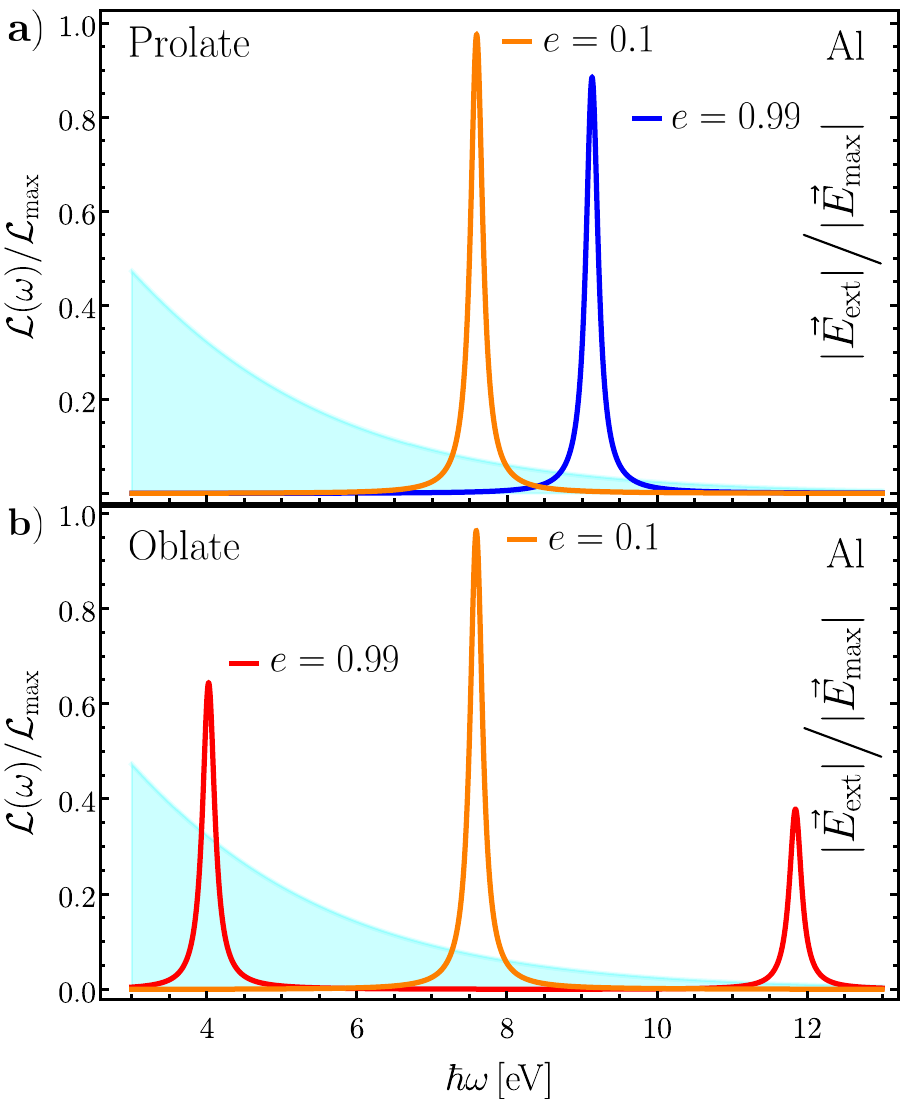}
    \caption{Normalized spectral density $\mathcal{L}$ for $\mathbf{a)}$ aluminum prolate and $\mathbf{b)}$ oblate ellipsoids with an eccentricity $e$ of 0.1 and 0.99, as a function of excitation energy $\hbar\omega$. Colors indicate the proximity of $e$ to $0$ (orange) or $1$ (blue for prolate, red for oblate). The cyan line corresponds to the spectral contribution of the external electric field, and the shaded area depicts the overlap with $\mathcal{L}$.}
    \label{fig:LEllipsoidsEccentricity}
\end{figure}

To study the relationship between momentum transfer and the eccentricity of the spheroidal NP, it is useful to analyze the spectral contributions across different eccentricity values. Figure \ref{fig:LEllipsoidsEccentricity} shows the dependence on eccentricity by exploring the spectral density $\mathcal{L}$ for prolate and oblate aluminum spheroids at two extreme eccentricity values: $e=0.1$ and $e=0.99$. A schematic representation of the magnitude of the external electric field generated by the swift electron is included in the figure in color cyan, with a normalized axis scale provided to the right. The cyan-shaded area below the external electric field allows the visualization of the overlap between the shifted plasmonic resonance of the NPs and the exciting external electromagnetic field. Notably, the overlap with redshifted resonances is significant due to the spectral contribution in lower energies of the electromagnetic field generated by the swift electron. Each subplot exhibits discernible peaks corresponding to plasmonic resonances, with distinct color coding for clarity.

For an eccentricity of $e=0.1$ (depicted in orange), the spheroidal nanoparticles closely approximate spherical shapes, resulting in resonance patterns similar to the spherical dipolar plasmonic resonance. As shown in Fig. \ref{fig:LEllipsoidsEccentricity}, in the extreme case of $e=0.99$, both prolate and oblate spheroidal NPs exhibit resonances, notably deviating from the spherical dipolar resonance. Note that as $e$ increases, the plasmonic resonance of the prolate NP\JLBG blueshifts [Fig. \ref{fig:LEllipsoidsEccentricity}$\mathbf{a)}$]. In contrast, the plasmonic resonance of the oblate NP splits into two resonances: one is blueshifted, and the other is redshifted, exhibiting a higher intensity [Fig. \ref{fig:LEllipsoidsEccentricity}$\mathbf{b)}$].

The insights from Figure \ref{fig:LEllipsoidsEccentricity} are useful to understand the behavior of the AMT as a function of the eccentricity observed in Figure \ref{fig: LToSphereEllipsoids}$\mathbf{c)}$. As eccentricity $e$ increases, the AMT to the prolate NP decreases compared to both the spherical and oblate NP [see Fig. \ref{fig: LToSphereEllipsoids}$\mathbf{c)}$]. This can be attributed to the reduced significance of the blueshifted plasmonic resonance in the interaction between the plasmonic resonance of the prolate NP and the electromagnetic field of the swift electron. The latter exhibits larger values at lower energies, as illustrated in Fig. \ref{fig:LEllipsoidsEccentricity} with the cyan-shaded area.

For low values of $e$, the AMT to the oblate NP is lower than that to the sphere. However, as $e$ increases, the AMT reaches a minimum. Beyond this threshold, a notable transition occurs, with the AMT to the oblate NPs surpassing that of the sphere [see Fig. \ref{fig: LToSphereEllipsoids}$\mathbf{c)}$]. This transition can be attributed to the growing influence of the interaction between the redshifted plasmonic resonance and the electromagnetic field of the swift electron. Despite the lower intensity of the redshifted resonance compared to the spherical plasmonic resonance (orange peak), as observed in Fig. \ref{fig:LEllipsoidsEccentricity}$\mathbf{b)}$, there is a more pronounced overlap between the spectral curve and the cyan shaded area, indicating a stronger interaction.

\begin{figure}[ht!]
    \centering
    \includegraphics[width=1\linewidth]{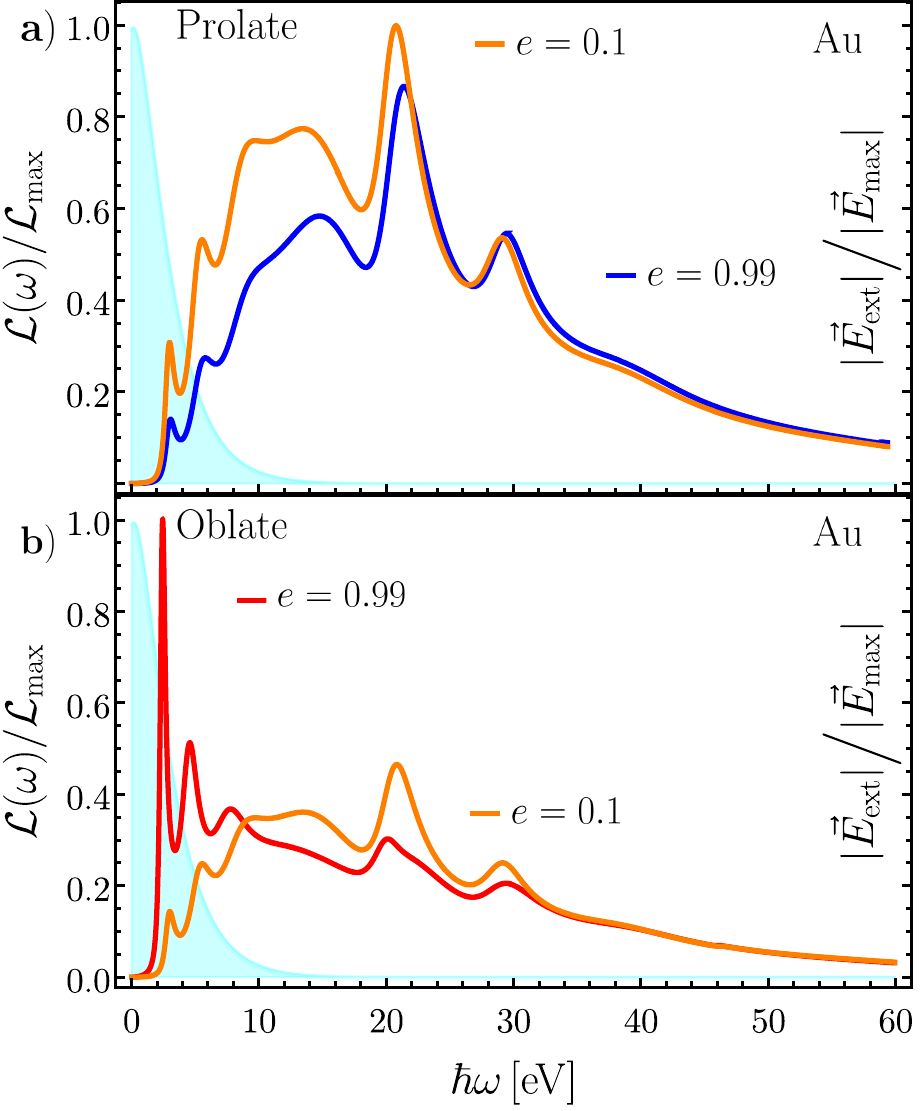}
    \caption{Normalized spectral density of AMT for gold nanospheroids as a function of excitation energy $\hbar \omega$, color-coded to indicate varying eccentricity $e$ values. In Panel $\mathbf{a)}$, ranging from orange (sphere-like NP, $e = 0.1$) to blue (prolate NP, $e=0.99$), and in Panel $\mathbf{b)}$, from orange (sphere-like NP, $e = 0.1$) to red (oblate NP, $e=0.99$). The Cyan shaded line is the spectral contribution of the external electric field, and the shaded area depicts the overlap with $\mathcal{L}$.}
    \label{fig: AMT Werner Au Spheroids}
\end{figure}

Interestingly, despite their different dielectric responses, small spheroidal NPs composed of Al and Au exhibit remarkably similar behaviors concerning AMT. This similarity can be appreciated by comparing the results of Al NPs and Au NPs presented in Appendix \ref{Appendix: Au NPs}. It is worth mentioning that the AMT to Au NPs exceeds that to Al NPs. This can be attributed to the broader distribution of plasmonic resonances within the dielectric function of gold across the energy spectrum. Furthermore, the significance of lower energies in shaping the spectral density $\mathcal{L}$ of Au NPs is accentuated, as depicted in Fig. \ref{fig: AMT Werner Au Spheroids}, where the Au spectral density of AMT is displayed. Different eccentricity values ($e=0.1$ in orange, $e=0.99$ in blue for prolate, and red for oblate) are color-coded to illustrate variations. The cyan curve represents the external electric field generated by the swift electron, aiding in visualizing the overlap of redshifted plasmonic resonances with the external electromagnetic field, akin to its role in Fig. \ref{fig:LEllipsoidsEccentricity}.

A thorough analysis of Fig. \ref{fig: AMT Werner Au Spheroids}$\mathbf{a)}$ reveals that, similar to Al prolate NPs shown in Fig. \ref{fig:LEllipsoidsEccentricity}$\mathbf{a)}$, the plasmonic resonances of Au prolate NPs undergo a blueshift. Additionally, at lower energies, where the external electromagnetic field becomes more intense, the intensities of Au prolate plasmonic resonances tend to diminish compared to the resonances of Au spherical NPs. Conversely, in the case of Au oblate NPs illustrated in Fig. \ref{fig: AMT Werner Au Spheroids}$\mathbf{b})$, the plasmonic resonances experience a redshift. Notably, at lower energy resonances, where the external electromagnetic field is more intense, there is an increase in intensity as $e$ increases. This behavior clarifies why the AMT for oblate spheroids surpasses that of spherical NPs.

\subsection{Angular momentum transfer for Platonic Solids}

In this section, we investigate the angular momentum transfer to Platonic Solids, a group of five geometric solids characterized by identical and regular polygonal faces. As discussed in Section \ref{sec: Platonic Solids}, these solids encompass the tetrahedron (or pyramid), hexahedron (or cube), octahedron, dodecahedron, and icosahedron (see Table \ref{tab: Platonic Solids}).

\begin{figure}[h!]
    \centering
    \includegraphics[width=1\linewidth]{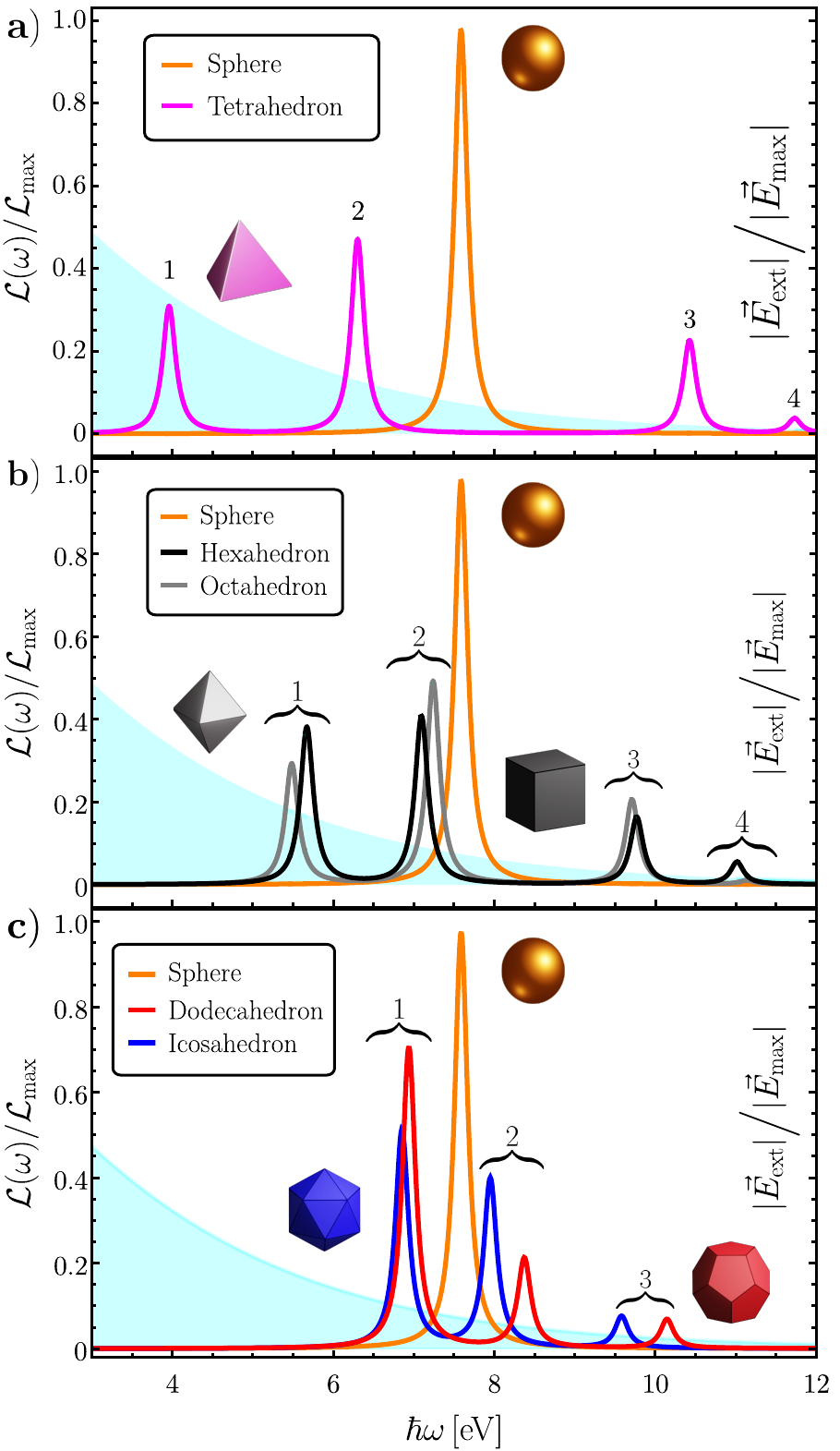}
    \caption{Normalized $\mathcal{L}(\omega)$ for reciprocal Platonic Solids characterized with aluminum modeled with the Drude model \cite{rakic}, for the sphere, $\mathbf{a)}$ the tetrahedron, $\mathbf{b)}$ the octahedron and hexahedron, and $\mathbf{c)}$ the icosahedron and dodecahedron. The resonances tend to group for reciprocal Platonic Solids, indicating similar AMT characteristics between these pairs. The cyan line corresponds to the spectral contribution of the external electric field, and the shaded area illustrates the overlap with $\mathcal{L}$.}
    \label{fig: SpectralDensities}
\end{figure}

For Al NPs, each Platonic Solid exhibits multiple dipolar resonances, which undergo significant shifts away from the spherical dipolar resonance, depending on their specific geometries. Notably, the plasmonic resonances of dual Platonic Solids\footnote{The tetrahedron is dual to itself, and the octahedron and hexahedron are dual to each other, as well as the icosahedron and the dodecahedron are dual to each other, as depicted in Table \ref{tab: Platonic Solids}.} tend to overlap. This phenomenon is depicted in Fig. \ref{fig: SpectralDensities}. Comparable shifted resonances are presented in the subfigures of Fig. \ref{fig: SpectralDensities}, sequentially labeled as 1, 2, 3, and 4. The cyan curve in Fig. \ref{fig: SpectralDensities} aids in highlighting the overlap between the external electric field and the redshifted plasmonic resonance of the NP, emphasizing the contribution of lower energies.

\begin{figure}[h!]
    \centering
    \includegraphics[width=1\linewidth]{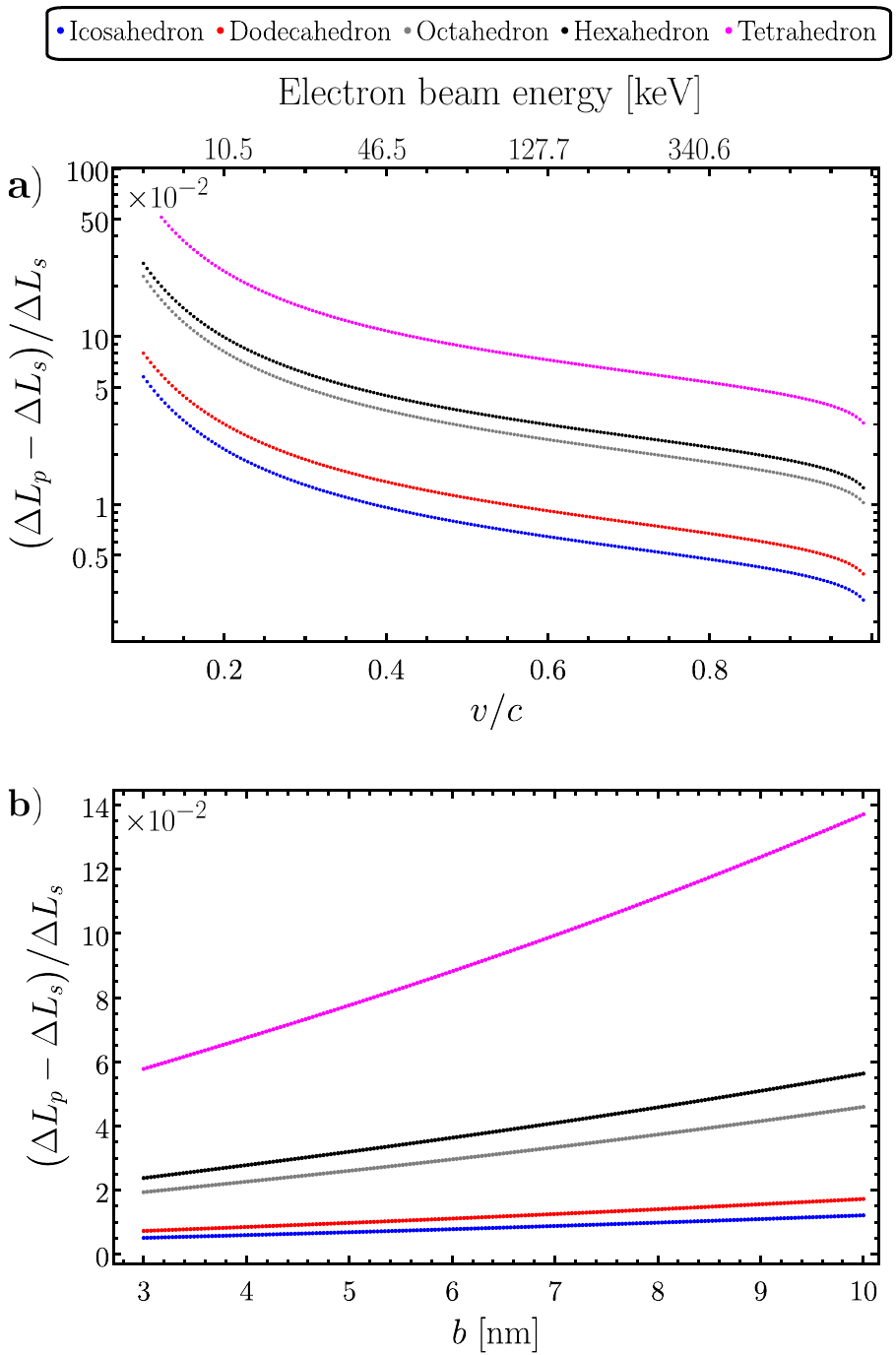}
    \caption{Relative difference in angular momentum transfer (AMT) to each Al Platonic Solid $\mathbf{a)}$ as a function of the electron speed $v/c$ with a fixed impact parameter $b = 3.5$ nm, and $\mathbf{b)}$ as a function of the impact parameter, with a fixed electron speed $v=0.7c$ ($200$ keV). }
    \label{fig: LToSphere}
\end{figure}

The tetrahedron exhibits the most pronounced redshift among all Platonic Solids, marked as resonance $1$ in Fig. \ref{fig: SpectralDensities}$\mathbf{a)}$, resulting in the highest AMT (see Fig. \ref{fig: LToSphere}). Additionally, the hexahedron and octahedron show multiple redshifted resonances, labeled as $1$ and $2$ in Fig. \ref{fig: SpectralDensities} $\mathbf{b)}$, respectively, while the dodecahedron and icosahedron redshift correspond only to resonance $1$ in Fig. \ref{fig: SpectralDensities} $\mathbf{c)}$. The higher intensity of the electromagnetic field produced by the electron at lower frequencies leads to an increased AMT with higher redshifts.

Based on the preceding discussion, we anticipate that the AMT to the tetrahedron will be the highest among all the Platonic Solids. Similarly, the AMT to the hexahedron and octahedron is expected to exceed that of the dodecahedron and icosahedron. However, distinguishing the larger AMT between the hexahedron and octahedron (duals) and between the dodecahedron and icosahedron (also duals) solely from the spectral contributions of their plasmonic resonances presents a challenge. Nonetheless, we anticipate that the dual particles will demonstrate very similar behavior, as their resonances are closely situated relative to each other.

We analyze the variation in AMT relative to a sphere for each Platonic Solid, composed of Al, as a function of the electron velocity $v/c$, with a constant impact parameter $b = 3.5 \text{nm}$, illustrated in Fig. \ref{fig: LToSphere}$\mathbf{a)}$\footnote{Fig. \ref{fig: LToSphere}$\mathbf{a)}$ utilizes a logarithmic scale in the vertical axis to aid in distinguishing the relative differences between dual Platonic Solids.}. Furthermore, Fig. \ref{fig: LToSphere}$\mathbf{b)}$ demonstrates the relative difference in AMT for each Platonic Solid as a function of the impact parameter while maintaining a constant electron speed of $v=0.7c$ ($200$ keV). Notably, every Platonic Solid exhibits a positive relative difference in AMT, indicating that the AMT to a Platonic Solid NP is consistently higher than that to spherical NPs with equivalent volumes. It is worth noting the trend for dual pairs of Platonic Solids to group together, in terms of relative AMT, as illustrated in Fig. \ref{fig: LToSphere}. In contrast, the relative difference in AMT for ellipsoids is negative, with only very high values of $e$ leading to the AMT to the oblate NP surpassing that to spherical NPs.

Similarly to the investigation conducted on ellipsoidal particles, the AMT to Au Platonic Solids NPs was explored. Despite the differences in dielectric functions between Al and Au, similar patterns of AMT were observed. This resemblance is seen when comparing Fig. \ref{fig: LToSphere} with Fig. \ref{fig:AMTPlatonicSolidsAuWerner} found in Appendix \ref{Appendix: Au NPs}. The latter is elucidated by analogous shifts of the resonances observed in Al Platonic Solids NPs, as depicted in Fig. \ref{fig:AMTPlatonicSolidsAuWerner}$\mathbf{a)}$.

%************************************************************
%************************************************************
%							CONCLUSIONS
%************************************************************
%************************************************************

\section{Conclusions}

We have presented an investigation of the angular momentum transfer to spheroidal and polyhedral (Platonic Solids) nanoparticles, from the electron beam inside a STEM with a classical electrodynamics approach in the small particle limit. The nanoparticles were assumed to be made of a homogeneous material. We modeled aluminum and gold nanoparticles, with gold nanoparticles demonstrating a higher angular momentum transfer for all geometries considered. The causality of the polarizabilities was revised to avoid non-physical angular momentum transfer predictions. We demonstrated that higher angular momentum transfer occurred as a result of redshifted resonances (compared to spherical nanoparticles of the same volume), attributed to the increased influence of lower-energy components within the electromagnetic field of the electron beam. Prolate nanoparticles showed a single blueshifted resonance leading to lower angular momentum transfer than spheres. However, oblate nanoparticles displayed two resonances, one blueshifted and the other redshifted with higher intensity. This led to a change in tendency, with angular momentum transfer surpassing that of spheres for high eccentricities. Platonic Solids exhibited multiple resonances, and sharper solids displayed greater redshift, resulting in higher angular momentum transfer. Remarkably, dual Platonic Solids had similar resonances between each other, leading to comparable angular momentum transfer. These findings highlight the potential to enhance and tune the magnitude of the angular momentum transfer, selecting geometries with sharper edges to promote peak effects.

%************************************************************
%************************************************************
%							ACKNOWLEDGMENTS
%************************************************************
%************************************************************
\section*{Acknowledgements}
 This work was partially supported by the UNAM-DGAPA PAPIIT IN107122 project. J. L. B.-G. acknowledges the support received from Conahcyt for CVU No. 1145427 under scholarship 799331. J. Á. C.-R. acknowledges the Swedish Research Council, Olle Engkvist’s Foundation, and Knut and Alice Wallenberg Foundation for financial support.
%
%%*******************************************************
%********************* Appendices **********************
%%*******************************************************
%
%\appendix
%

\begin{appendices}

%
%
%%*******************************************************
%********************* Appendix A **********************
%%*******************************************************
%
\section{Components of the spectral density: complete formulas}\label{Appendix: Spectral densities}
\renewcommand{\theequation}{A.\arabic{equation}}
\setcounter{equation}{0}

The formulas for the components of the spectral density of the angular momentum transfer from a swift electron to a non-spherical NP in the dipole approximation were derived in Ref. \cite{castellanos2021angular}. As stated in the main text, the angular momentum transfer (AMT) can be calculated in terms of $\tensa{\alpha}$ and $\omega$ as follows:
\begin{equation}
\Delta \vv{L} = \int_0^{\infty} \vvb{\mathcal{L}}(\tensa{\alpha}; \omega) d\omega,
\label{eq: DL integral spectral density 2}
\end{equation}
where the spectral density can be written in terms of its cartesian components
\begin{align}
    \mathcal{L}_{\rm x}\var{\tensa{\alpha}; \w}=& \var{\frac{1}{4\pi\epsilon_0}}^2 \frac{4 q_e^2}{v^4 \gamma^2}\frac{\epsilon_0}{\pi} \bigg[\frac{\omega^2}{\gamma^2} K_0^2\wbgv \times \nonumber\\
    & \text{Re} \Big\{ \alpha_{\rm yz}\var{\w} \Big\} + \frac{\abs{\w}\w}{\gamma} K_0\wbgv \times \nonumber\\
    & K_1\wbgv \text{Im} \Big\{ \alpha_{\rm yx}\var{\w} \Big\}\bigg], \\
     \mathcal{L}_{\rm y}\var{\tensa{\alpha}; \w}=& \var{\frac{1}{4\pi\epsilon_0}}^2 \frac{4 q_e^2}{v^4 \gamma^2}\frac{\epsilon_0}{\pi} \bigg[\omega^2 K_1^2\wbgv \times \nonumber\\
    & \text{Re} \Big\{ \alpha_{\rm zx}\var{\w} \Big\} - \frac{\w^2}{\gamma^2} K_0^2\wbgv \times \nonumber\\
    & \text{Re} \Big\{ \alpha_{\rm xz}\var{\w} \Big\} - \frac{\abs{\w}\w}{\gamma}  K_0\wbgv \times \nonumber\\
    & K_1\wbgv \text{Im} \Big\{ \alpha_{\rm xx}\var{\w} + \alpha_{\rm zz}\var{\w} \Big\}\bigg],\\
    \mathcal{L}_{\rm z}\var{\tensa{\alpha}; \w}=& \var{\frac{1}{4\pi\epsilon_0}}^2 \frac{4 q_e^2}{v^4 \gamma^2}\frac{\epsilon_0}{\pi} \bigg[\frac{\abs{\w} \w}{\gamma} K_0\wbgv \times \nonumber\\
    & K_1\wbgv \text{Im} \Big\{ \alpha_{\rm yz}\var{\w} \Big\} - \nonumber \\
    &\w^2 K_1^2\wbgv \text{Re}\Big\{ \alpha_{\rm yx}\var{\w} \Big\}\bigg],
\end{align}
where $\alpha_{\text{ij}}$ is the $\text {ij}$ component of the electric polarizability tensor, $-q_e$ is the electric charge of the electron, $K_{\nu}$ is the modified Bessel function of the second kind of order $\nu$, $v$ is the constant electron speed, $\gamma = (1-\beta^2)^{-1/2}$ is the Lorentz factor, with $\beta = v/c$, $c$ is the speed of light, and $\text{Re} \{z\}$, $\text{Im} \{z\}$ denote the real and imaginary parts of $z$, respectively.
%
%
%%*******************************************************
%********************* Appendix B **********************
%%*******************************************************
%
\section{Causality analysis and geometric factor calculation for the polarizabilities of spheroidal nanoparticles}\label{Appendix: spheroids}
\renewcommand{\theequation}{C.\arabic{equation}}
\setcounter{equation}{0}

\begin{figure}[h!]
\centering
\includegraphics[width=1\linewidth]{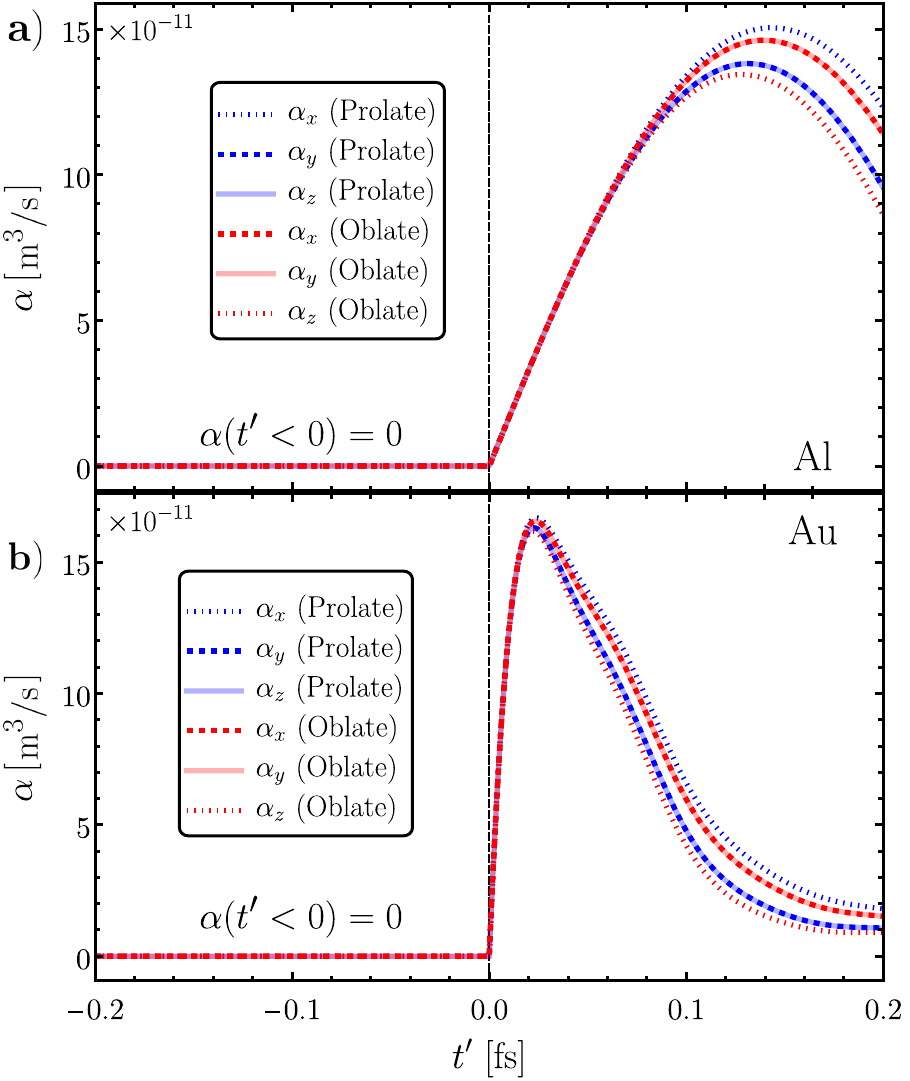}
\caption{\label{fig: polCausalityEllipsoids} 
Polarizabilities as a function of time for different directions of the oblate and prolate response considering two materials: $\mathbf{a)}$ aluminum modeled with the Drude model \cite{rakic}, and $\mathbf{b)}$ gold from Werner et al. \cite{werner}. The polarizabilities exhibit a zero response for $t^{\prime} < 0$, indicating a causal behavior of each system.}
\end{figure}

The causality analysis involves computing the time-dependent polarizability of oblate and prolate spheroidal nanoparticles with volume $V=4\pi a^3/3$ and $a=1 \text{nm}$, as illustrated in Fig. \ref{fig: polCausalityEllipsoids}. The blue lines represent the three different cartesian directions of the polarizability of the prolate NP, while the red lines correspond to the three different directions of the polarizability of the oblate NP (see Table~\ref{tab: Ellipsoids}). Each direction was plotted with a different line style (continuous, dashed, and dotted) to aid in differentiating between the three different directions of the same ellipsoidal NP. We considered the Drude-model dielectric function or Ref. \cite{rakic} for the aluminum NPs in Fig. \ref{fig: polCausalityEllipsoids}$\mathbf{a)}$  and the the Drude-Lorentz model of Ref. \cite{werner} for Fig. \ref{fig: polCausalityEllipsoids}$\mathbf{b)}$.

The polarizability as a function of time is calculated through a Fourier Transform of Eq. \eqref{eq: alpha_spheroids}. The results reveal that the response for times \(t^{\prime} < 0\) is zero, demonstrating the causal behavior of each system. It is worth mentioning that \(t^{\prime}\) is the response time of the electromagnetic excitations within the NP, which is different from the time \(t\) related to the electron's movement.

The calculations of the geometrical factors appearing in Eq. \eqref{eq: alpha_spheroids} are reproduced in this Appendix from  Ref. \cite{bohren}. The surface of the ellipsoids is defined by the equation
\begin{equation*}
\frac{x^2}{a_x^2}+\frac{y^2}{a_y^2}+\frac{z^2}{a_z^2}=1,
\end{equation*}
where $a_x$, $a_y$, and $a_z$ are the semi-axes with $a_z \le a_x \le a_y$. The geometric factors $L_i$ in Eq. \eqref{eq: alpha_spheroids} are given by
\begin{equation*}
L_i = \frac{a_x a_y a_z}{2} \int_0^{\infty} \frac{dq}{(a_i^2+q) f(q)},
\end{equation*}
with $f(q) = \sqrt{(q+a_x^2)(q+a_y^2)(q+a_z^2)}$.
However, only two out of the three geometric factors $L_x$, $L_y$, and $L_z$ are independent, due to the relation
\begin{equation*}
L_x + L_y + L_z = - a_x a_y a_z \int_0^{\infty}\frac{d}{dq} \frac{1}{f(q)} \, dq = 1.
\end{equation*}

For a prolate spheroid, where $a_x = a_z$ and $L_x = L_z = (1-L_y)/2$, and the expression for $L_y$ is given by 
\begin{equation*}
L_y = \frac{1-e^2}{e^2} \left[-1 + \frac{1}{2e} \ln \left(\frac{1+e}{1-e}\right)\right],
\end{equation*}
where $e^2 = 1 - (a_x^2 / a_y^2)$ corresponds to the eccentricity of the prolate spheroid.

For an oblate spheroid, where $a_x = a_y$ and $L_x = L_y = (1-L_z)/2$, and the expression for $L_y$ is 
\begin{align*}
L_y =& \frac{1}{2e^2}\sqrt{\frac{1-e^2}{e^2}} \left[ \frac{\pi}{2} - \arctan \left(\sqrt{\frac{1-e^2}{e^2}} \right) \right] - \frac{1-e^2}{2e^2} ,
\end{align*}
where $e^2 = 1 - (a_z^2 / a_y^2)$ represents the eccentricity of the oblate spheroid.
%
%
%%*******************************************************
%********************* Appendix C **********************
%%*******************************************************
%
\section{Polarizabilities of Platonic Solids }\label{Appendix: Polarizabilities}
\renewcommand{\theequation}{B.\arabic{equation}}
\setcounter{equation}{0}

In reference \cite{sihvola2004polarizabilities}, the polarizabilities of Platonic Solids (see Table \ref{tab: Platonic Solids}) were calculated using numerical methods. These polarizabilities were then approximated via analytical formulas employing Padé approximants, which consists of a quotient of polynomials. The choice of  Padé approximants offers several advantages. First, as stated in \cite{sihvola2004polarizabilities}, compared to a Taylor series approximation, the numerical convergence of the  Padé approximants is significantly faster. Additionally, when calculating the poles of the Padé approximants, they are found to be related to plasmonic resonances, which represent crucial spectral properties of the nanoparticles. Lastly, causality is challenging to ensure using only a Taylor approximation, thus such Taylor expression may lead to non-physical momentum transfer calculations.

In this Appendix, we reproduce the results published in \cite{sihvola2004polarizabilities} of the normalized polarizability $\alpha_n = \alpha\big/V$, where $V$ represents the volume of the particle. The formulas are presented in the form of Eq. \eqref{eq: polarizability PS}, where $\alpha(\omega) \sim P(\tilde{\epsilon})\big/Q(\tilde{\epsilon})$, and for each particle the polynomials $P(\tilde{\epsilon})$ and $Q(\tilde{\epsilon})$ are given below:

\subsection{Tetrahedron}
\vspace{-0.3cm}
\begin{align*}
P(\tilde{\epsilon})&=5.0285 (\tilde{\epsilon}^3+7.65667\tilde{\epsilon}^2+8.50919\tilde{\epsilon}+1.8063) \\
Q(\tilde{\tilde{\epsilon}})&=\tilde{\epsilon}^4+14.1983\tilde{\epsilon}^3+44.9182\tilde{\epsilon}^2+30.2668\tilde{\epsilon}+5.0285
\end{align*}

\subsection{Cube}
\vspace{-0.3cm}
\begin{align*}
P(\tilde{\epsilon})&=3.6442 (\tilde{\epsilon}^3+4.83981\tilde{\epsilon}^2+5.54742\tilde{\epsilon}+1.6383) \\
Q(\tilde{\epsilon})&=\tilde{\epsilon}^4+8.0341\tilde{\epsilon}^3+19.3534\tilde{\epsilon}^2+15.4349\tilde{\epsilon}+3.6442
\end{align*}

\subsection{Octahedron}
\vspace{-0.3cm}
\begin{align*}
P(\tilde{\epsilon})&=3.5507 (\tilde{\epsilon}^3+5.13936\tilde{\epsilon}^2+5.86506\tilde{\epsilon}+1.5871) \\
Q(\tilde{\epsilon})&=\tilde{\epsilon}^4+8.26227\tilde{\epsilon}^3+19.8267\tilde{\epsilon}^2+15.6191\tilde{\epsilon}+3.5507
\end{align*}

\subsection{Dodecahedon}
\vspace{-0.3cm}
\begin{align*}
P(\tilde{\epsilon})&=3.1779 (\tilde{\epsilon}^2+2.42101\tilde{\epsilon}+1.24643) \\
Q(\tilde{\epsilon})&=\tilde{\epsilon}^3+4.72932\tilde{\epsilon}^2+6.53464\tilde{\epsilon}+2.56842
\end{align*}

\subsection{Icosahedron}
\vspace{-0.3cm}
\begin{align*}
P(\tilde{\epsilon})&=3.1304 (\tilde{\epsilon}^2+3.04968\tilde{\epsilon}+1.99016) \\
Q(\tilde{\epsilon})&=\tilde{\epsilon}^3+5.29169\tilde{\epsilon}^2+8.52687\tilde{\epsilon}+4.0896
\end{align*}

%
%
%               APPENDIX: EXTERNAL ELECTRIC FIELD
%
%
\section{Electromagnetic field produced by a swift electron}\label{Appendix: Ext Field}
\renewcommand{\theequation}{D.\arabic{equation}}
\setcounter{equation}{0}

\begin{figure}[h!]
    \centering
    \includegraphics[width=\linewidth]{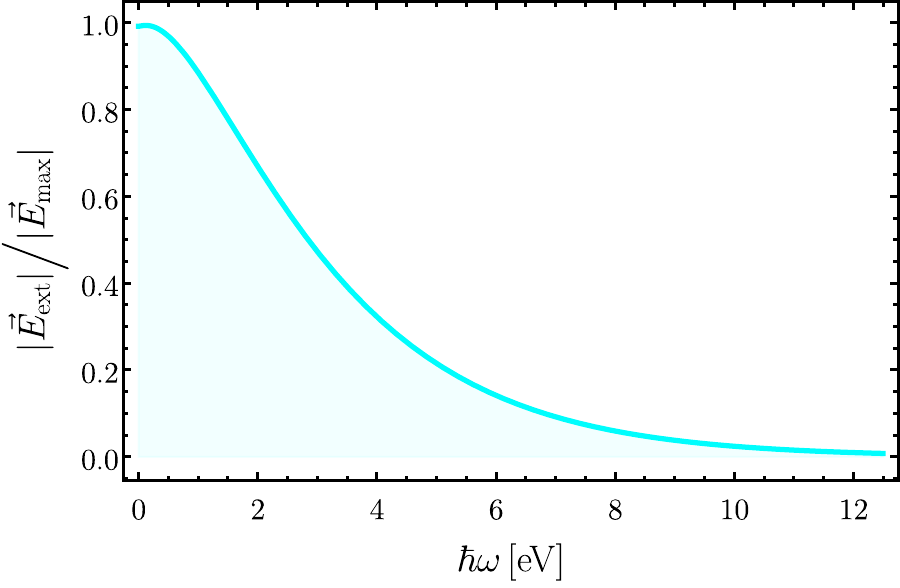}
    \caption{Magnitude of the external electric field vs. energy $\hbar \omega$, emphasizing the higher contribution from lower frequencies.}
    \label{fig: ext EM field}
\end{figure}

Throughout this work, we discuss the spectral characteristics of the electromagnetic field generated by a swift electron. This electromagnetic field serves as a source of evanescent radiation spanning the entire frequency spectrum, having its most significant contribution at lower frequencies, as illustrated in Fig. \ref{fig: ext EM field}. This is important when discussing geometries that enhance angular momentum transfer, as we explore in this work. Specifically, those geometries that result in shifts of the plasmonic resonances to these lower frequencies. A detailed study about this electromagnetic field can be consulted in Ref.~\cite{maciel2019electromagnetic}.

\section{Response of Au nanoparticles}\label{Appendix: Au NPs}
\renewcommand{\theequation}{D.\arabic{equation}}
\setcounter{equation}{0}

\begin{figure}[h!]
    \centering
    \includegraphics[width=1\linewidth]{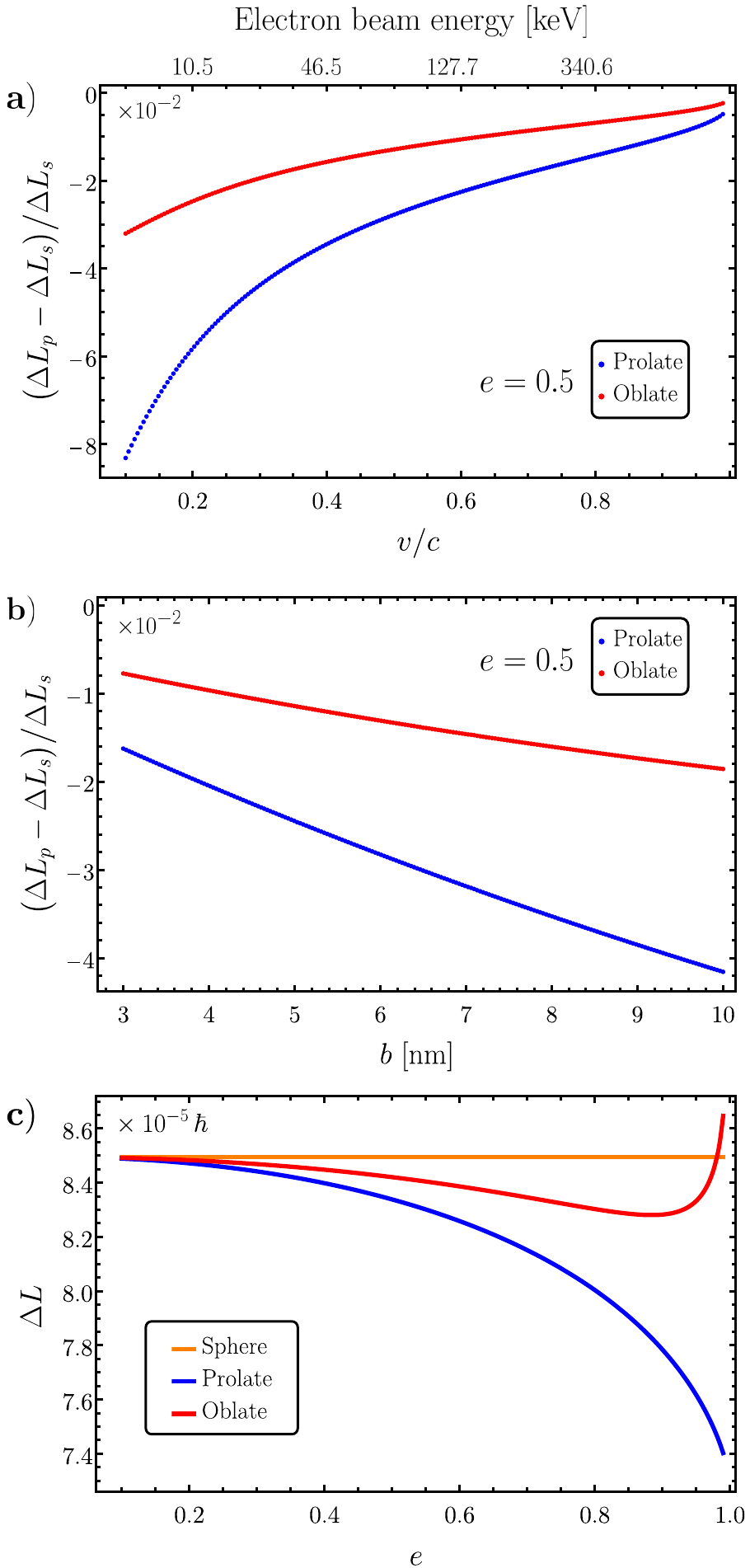}
    \caption{Relative difference in angular momentum transfer to a gold ellipsoidal nanoparticle as a function of $\mathbf{a)}$ the electron's speed $v$ with impact parameter $b = 3.5 \text{nm}$, and $\mathbf{b)}$ the impact parameter $b$ with $v = 0.7c$ ($200$ keV), for both prolate (blue) and oblate (red) ellipsoids. The eccentricity of the ellipsoids was set to $e = 0.5$. $\mathbf{c)}$ Angular momentum transfer to an small ellipsoidal gold NP as a function of the eccentricity, with $b = 3.5 \text{nm}$ and $v = 0.7c$ ($200$ keV).}
    \label{fig: LToSphereEllipsoids-WernerAu}
\end{figure}

The comparison of the angular momentum transfer to Au and Al NPs yields intriguing insights. Despite the variations in the spectral structure of the momentum spectral density ($\mathcal{L}$), the AMT to Au NPs presents a remarkable resemblance to that of Al NPs in the spheroidal case. The dependency of AMT on parameters such as beam energy, impact parameter, and eccentricity exhibits a consistent functional form between Au and Al NPs, as can be concluded by comparing Figures \ref{fig: LToSphereEllipsoids} and \ref{fig: LToSphereEllipsoids-WernerAu}. Notably, both sets of NPs display analogous behavior in terms of AMT tunability, which depends on the geometry of the small ellipsoidal NP. However, it is noteworthy that the AMT to Au NPs consistently surpasses that to Al NPs.

Furthermore, a closer examination of the AMT to prolate and oblate NPs sheds light on other intriguing trends. Specifically, the AMT to prolate NPs decreases as eccentricity increases, while the AMT to oblate NPs follows a different trend, surpassing the AMT to a spherical NP beyond a certain eccentricity threshold. These behaviors are elucidated in Fig. \ref{fig: LToSphereEllipsoids-WernerAu} $\mathbf{c)}$.

\begin{figure}[h!]
    \centering
    \includegraphics[width=1\linewidth]{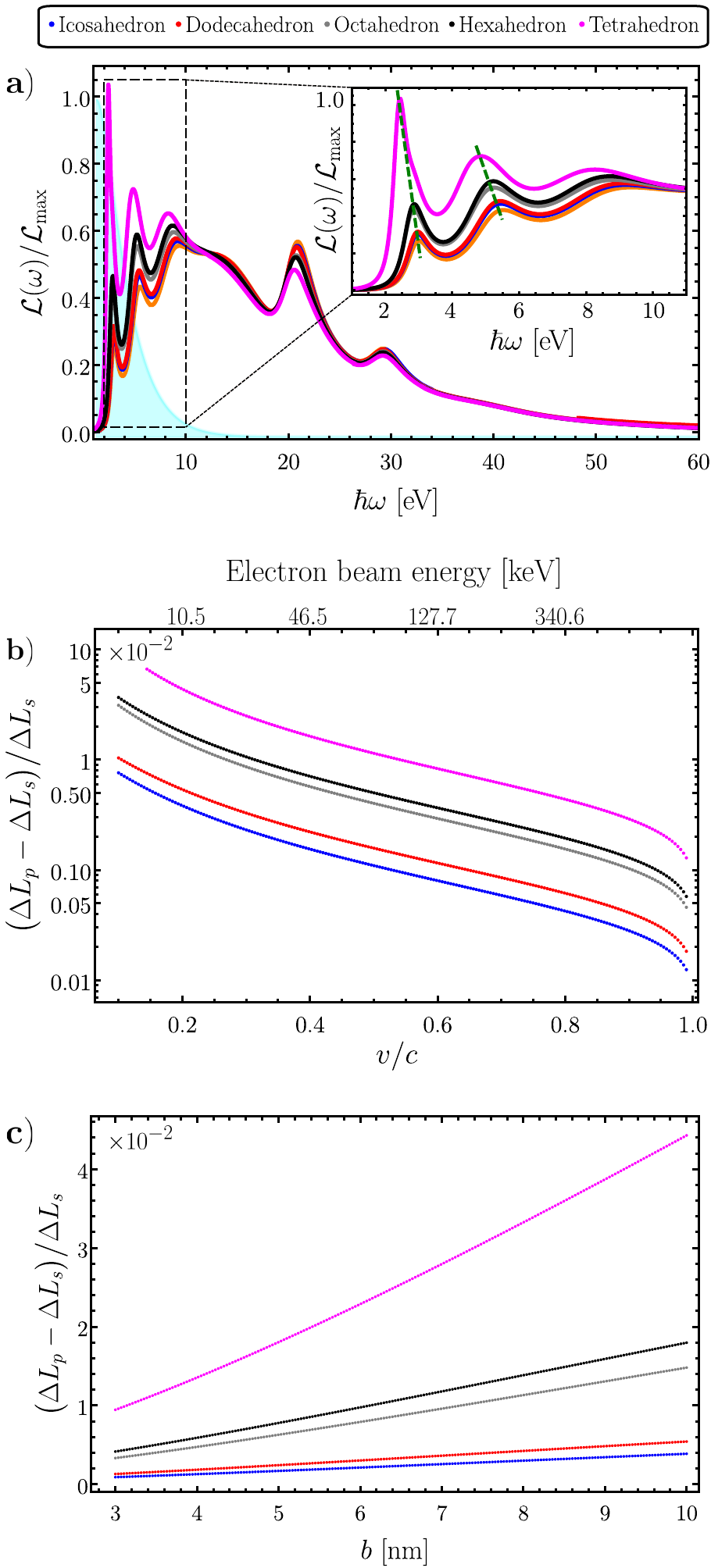}
    \caption{$\mathbf{a)}$ Spectral densities of angular momentum transfer for gold nanoparticles shaped as Platonic Solids. The inset provides a zoomed-in view of the low-frequency region, emphasizing the redshift (see green-dashed line). The relative difference in angular momentum transfer to gold polyhedral nanoparticles, $\mathbf{b)}$ as a function of the electron beam's speed with a constant impact parameter $b = 3.5 \, \text{nm}$, and $\mathbf{c)}$ as a function of the impact parameter with a constant electron speed $v= 0.7 \, c$ ($200$ keV). The dual pair of solid responses tend to group, with sharper polyhedral shapes exhibiting higher relative differences in angular momentum transfer values.}
    \label{fig:AMTPlatonicSolidsAuWerner}
\end{figure}

When analyzing the spectral density utilizing the dielectric function of gold NPs for Platonic Solids, an intricate pattern emerges, characterized by multiple resonances distributed throughout the energy spectrum. Despite this complexity, the spectral density of Au NPs consistently displays prominent redshifts, as illustrated in Fig. \ref{fig:AMTPlatonicSolidsAuWerner}$\mathbf{a)}$. This is why NPs with sharper polyhedral shapes tend to exhibit higher values of AMT. In Fig. \ref{fig:AMTPlatonicSolidsAuWerner}$\mathbf{a)}$, the inset provides a close-up view of the low-frequency region, accentuating the redshift effect for enhanced clarity (see green-dashed lines). It is also worth mentioning that resonances associated with reciprocal Platonic Solids tend to overlap, resulting in similar AMT values.

Moreover, the spectral densities for aluminum Platonic Solid nanoparticles present the same pattern: as the edges of the solids become sharper, the resonances exhibit a greater degree of redshift, accompanied by a heightened spectral intensity of the redshifted peaks. A comparison between Fig. \ref{fig: LToSphere} with Figs. \ref{fig:AMTPlatonicSolidsAuWerner}$\mathbf{b)}$ and \ref{fig:AMTPlatonicSolidsAuWerner}$\mathbf{c)}$ reveals a remarkable resemblance in the AMT to Al and Au polyhedral NPs. Furthermore, the response of reciprocal solids tends to cluster together, with the sharpest polyhedral shapes displaying the highest AMT values. In Fig. \ref{fig:AMTPlatonicSolidsAuWerner}$\mathbf{b)}$, the relative difference in AMT is shown as a function of the electron beam's speed, with a constant impact parameter of $b = 3.5 \, \text{nm}$. Conversely, in Fig. \ref{fig:AMTPlatonicSolidsAuWerner}$\mathbf{c)}$, the relative difference in AMT is shown as a function of the impact parameter, with a constant electron speed of $v= 0.7 \, c$ ($200$ keV). The AMT to Au NPs surpasses that of Al NPs due to the broader distribution of plasmonic resonances in the frequency domain and the significant contribution of lower frequencies to the spectral density of AMT.

\end{appendices}

%% ----------Loading bibliography style file------------
%\bibliographystyle{model1-num-names} %not included in template
%\bibliographystyle{cas-model2-names} %original from template
%%----------------------
%-----others-----------
%%  taken from https://support.stmdocs.in/wiki/index.php?title=Model-wise_bibliographic_style_files
%
%
%I was redirected here by https://support.stmdocs.in/wiki/index.php?title=Elsarticle_-_CAS#Bibliography 
%%
%\bibliographystyle{model1a-num-names} % cool but no title
%\bibliographystyle{model3-num-names} % includes title but don't love the order
\bibliographystyle{elsarticle-num-names}
\bibliography{0-references}

\end{document}